\documentclass{article}
\usepackage{spconf,amsmath,graphicx}
\usepackage{xcolor}
\usepackage{amsmath,bbm}
\usepackage{amssymb}
\usepackage{booktabs}
\usepackage{multirow}
\usepackage{cite}
\usepackage[hidelinks]{hyperref}
\usepackage[mathscr]{euscript}
\usepackage{fancyhdr}
\usepackage{xcolor}
\usepackage{subfig}
\usepackage{enumitem}
\usepackage{flushend}

\usepackage{fancyhdr}
\fancypagestyle{firstpage}{
  \fancyhf{}
  \setlength{\headheight}{20mm}
  \renewcommand{\topmargin}{-15mm}
  \fancyhead[C]{\vspace{5mm} \small \textit{Proc. IEEE ICIP 2023}, Kuala Lumpur, Malaysia, Oct. 2023\\ © 2023 IEEE. Personal use of this material is permitted. Permission from IEEE must be obtained for all other uses, in any current or future media, including reprinting/republishing this material for advertising or promotional purposes, creating new collective works, for resale or redistribution to servers or lists, or reuse of any copyrighted component of this work in other works.}}

\graphicspath{ {./figures/} }

\DeclareMathAlphabet\mathbfcal{OMS}{cmsy}{b}{n}

\title{Grad-FEC: Unequal Loss Protection of Deep Features in Collaborative Intelligence}
\name{Korcan Uyanik, S. Faegheh Yeganli, and Ivan V. Baji\'{c}\thanks{This work was supported in part by the NSERC grants RGPIN-2021-02485 and RGPAS-2021-00038.}}
\address{School of Engineering Science, Simon Fraser University, Burnaby, BC, Canada}
\begin{document}
%
\maketitle


%
\begin{abstract}
Collaborative intelligence (CI) involves dividing an artificial intelligence (AI) model into two parts: front-end, to be deployed on an edge device, and back-end, to be deployed in the cloud. The deep feature tensors produced by the  front-end are transmitted to the cloud through a communication channel, which may be subject to packet loss.  To address this issue, 
in this paper, we propose a novel approach to enhance the resilience of the CI system in the presence of packet loss through Unequal Loss Protection (ULP). The proposed ULP approach involves a feature importance estimator, which estimates the importance of feature packets produced by the front-end, and then selectively applies Forward Error Correction (FEC) codes to protect important packets. Experimental results demonstrate that the proposed approach can significantly improve the reliability and robustness of the CI system in the presence of packet loss. 
\end{abstract}


\thispagestyle{firstpage}

\section{Introduction}
\label{sec:intro}

The deployment of the Internet of Things (IoT) infrastructure offers numerous opportunities for innovative applications that rely on deep neural networks (DNNs) to process the acquired sensor data. 
However, due to the limited computational and energy resources of edge devices, researchers are investigating ways in which DNN-based analysis of sensory signals acquired at the edge can be most effectively realized. One of the promising strategies is 
Collaborative Intelligence (CI)~\cite{kang2017neurosurgeon,CI_overview_ICASSP2021}, which utilizes both edge and cloud resources to enhance the speed and efficiency of DNN computing~\cite{intelligent_sensing_TIP2020}.
Typically, CI involves dividing the DNN between an edge device and the cloud, where the edge device runs the DNN front-end and computes features that are then sent to the cloud to be processed by the DNN back-end. 
Due to the imperfections of real communication channels, transmitting the intermediate feature tensor can result in bit errors at the physical layer and subsequent packet loss at the transport or application layer~\cite{CI_overview_ICASSP2021}. As a result, 
error/loss resilience strategies should be incorporated into CI system design to enable effective and accurate analysis of the sensed signals.

The error/loss resilience in CI is a  
relatively unexplored topic.
Joint source-channel coding of deep features for the simple binary symmetric channel and the binary erasure channel has been considered in~\cite{NJSCC_ICML2019,BottleNet++}. For transmission over packet networks, several loss concealment approaches have been proposed in~\cite{ALTeC,inpainting_ICC2021,CALTeC}.
However, there is still limited understanding of the relevant trade-offs and a lack of design guidelines for error/loss resilience in CI. 

In this paper, we propose a novel approach to improve the resilience of a CI system to packet loss by using Unequal Loss Protection (ULP) for deep feature transmission. 
A crucial question to answer in this context is -- \emph{how can we tell which features are more important than others}? To answer this question, we employ a well-known method from the domain of explainable AI, namely Gradient-weighted Class Activation Mapping (Grad-CAM)~\cite{Grad-CAM}. For a given input image, Grad-CAM estimates the importance of each pixel according to its contribution to making the correct inference decision. We adjust Grad-CAM to estimate the importance of features being transmitted, rather than the input. However, even such a modified Grad-CAM is not realizable in the context of CI, because the edge device -- where these estimates need to be made -- does not have access to the inference decision. Hence, we develop a proxy model for Grad-CAM, which can approximate this estimate by observing only the input image. 
In summary, our contributions are as follows: \vspace{-3pt}
\begin{itemize}[leftmargin=*]
    \item We demonstrate that a Grad-CAM-like approach can reliably estimate the importance of features being transmitted in a CI system. \vspace{-3pt}
    \item We develop a model that can approximate Grad-CAM estimates without access to the inference decision. \vspace{-3pt}
    \item We show that a ULP approach based on such importance estimates is capable of providing significant loss resilience to a CI system based on ResNet-50~\cite{Resnet}. \vspace{-3pt}
\end{itemize}

The paper is organized as follows. Section ~\ref{sec:related_work}  briefly reviews
the related work on loss/error resilience in CI. The proposed methods -- feature importance estimation and unequal loss protection -- are described in Section~\ref{sec:propoon sed}. The experimental results are presented in Section~\ref{sec:experiments}, followed by the conclusions in Section~\ref{sec:conclusion}.

\section{Related work}
\label{sec:related_work}
Choi \emph{et al.}~\cite{NJSCC_ICML2019} 
developed a neural network for joint source-channel coding of intermediate features for discrete channels based on the maximization of mutual information between the image source data and the noisy latent codeword. 
Specifically, in this approach,
the binary symmetric channel (BSC) and binary erasure channel (BEC)  
are considered.
Another work~\cite{BottleNet++} designed an end-to-end trainable architecture called BottleNet++ for compressing and transmitting DNN features over a BEC or an Additive White Gaussian Noise (AWGN) channel. 
It is noted that the aforementioned approaches are targeted at improving feature transmission robustness against bit errors; in other words, they are physical-layer techniques. 

Often, physical-layer bit errors manifest themselves as packet loss at the application layer. 
In~\cite{ALTeC}, well-known tensor completion methods, 
namely simple low-rank tensor completion (SiLRTC), and highly accurate low-rank tensor completion (HaLRTC), were used to recover missing data in the deep feature tensor. Moreover, an approach called adaptive linear tensor completion (ALTeC) was developed, which was  
much faster and as accurate as SiLRTC and HaLRTC, but it required pre-training for a specific DNN backbone. 
In~\cite{CALTeC}, another approach called content adaptive linear tensor completion (CALTeC) was developed based on estimating a linear relationship between missing and available features, which did not require pre-training for a specific DNN backbone. Another backbone-agnostic approach for missing feature recovery was developed in~\cite{inpainting_ICC2021} using the concept of inpainting.

Our focus on this paper is on ULP which, to our knowledge, has not been studied in the context of CI. However, ULP was a popular topic in image and video communication research~\cite{ULP_Mohr_JSAC2000,ULP_vdS_TMM2001,ULP_SPIC2003}. In that line of research, a core question was how to decide which parts of the image/video are more important than others, so that they could be adequately protected. In this paper, we ask the same question for deep features being transmitted, and we offer a way to answer it.

\section{Proposed method}

\subsection{System overview}

\begin{figure}[t]
      \centering
      \includegraphics  [width=0.49\textwidth]{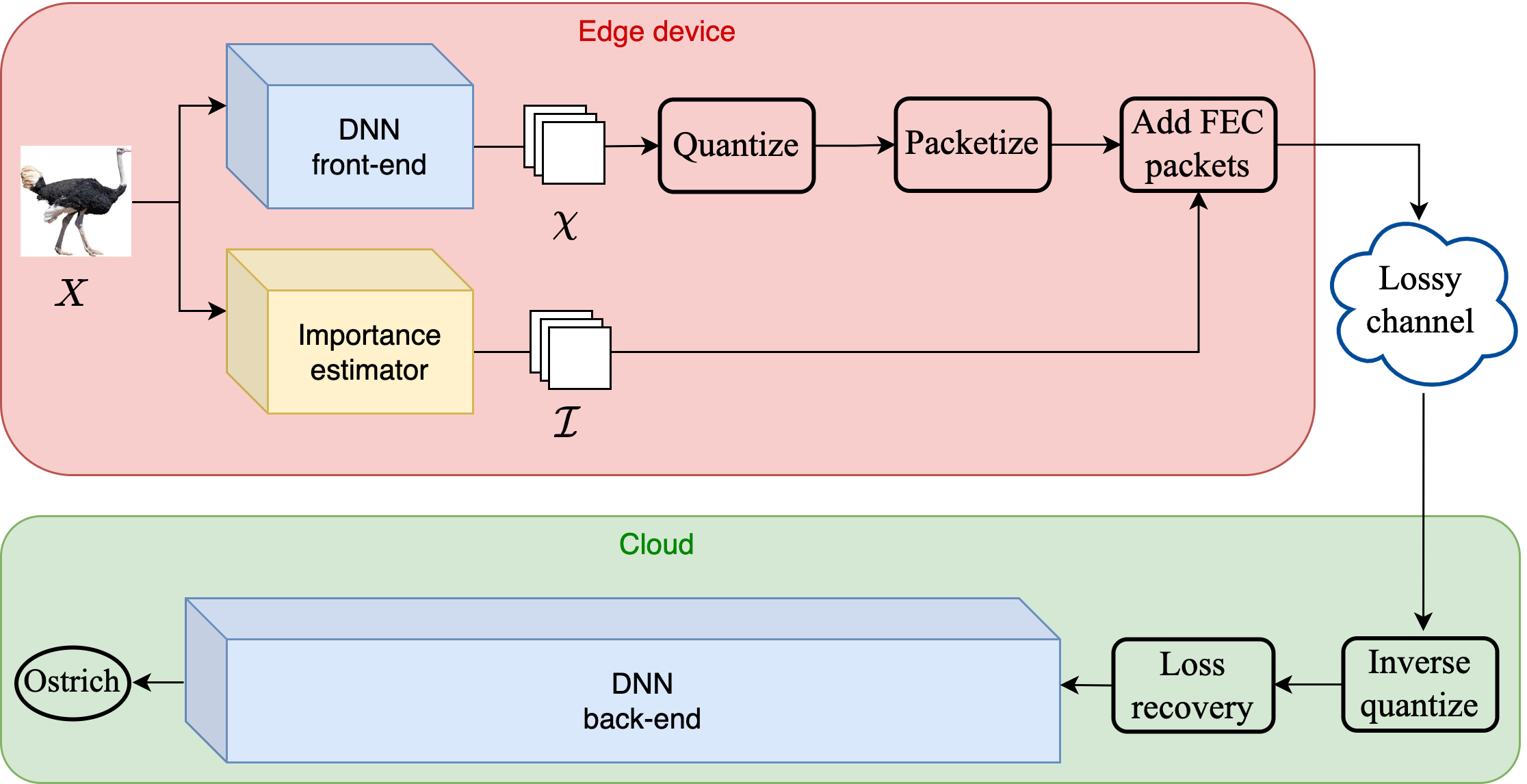}
      \caption{System overview.}
      \label{Figure 1}
\end{figure}

\begin{figure}[t]
    \centering
    \includegraphics[width=0.33\textwidth]{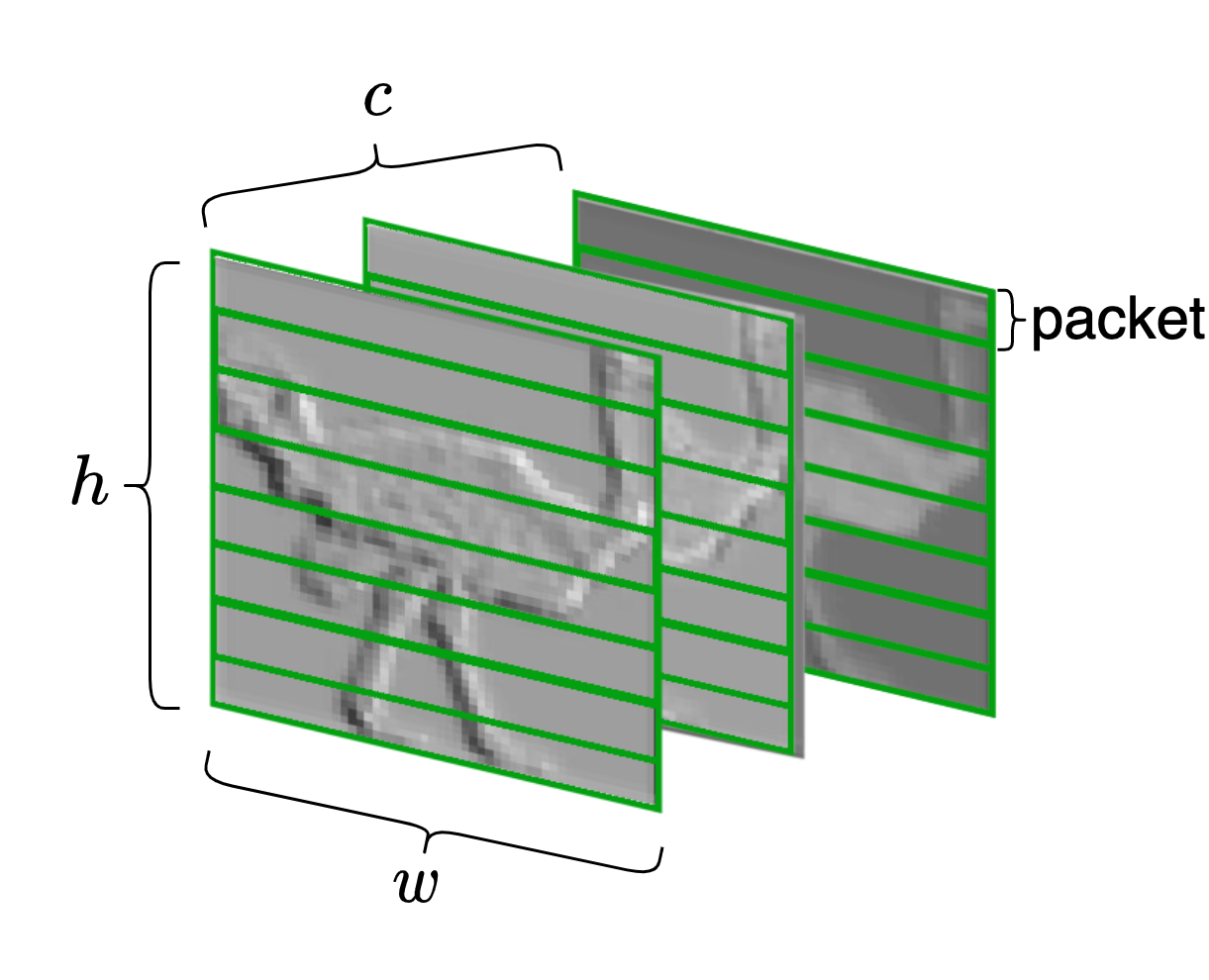}
    \vspace{-10pt}
    \caption{A feature tensor $\chi \in \mathbb{R}^{h\times w \times c}$ divided into packets.}
    \label{fig:tensor_packets}
\end{figure}

The proposed approach is depicted in Fig.~\ref{Figure 1}. 
The objective is to mitigate the impact of packet loss during feature transmission from the edge device to the cloud over a packet loss channel by using Forward Error Correction (FEC) to provide ULP.  
The edge device processes the input image $X$ and generates a deep feature tensor $\chi \in \mathbb{R}^{h\times w \times c}$, where $h$, $w$, and $c$ represent the width, height, and the number of tensor channels, respectively. 
The deep feature tensor is then 8-bit quantized and packetized for transmission over the packet network. We adopt the packetization scheme from~\cite{CALTeC}, where $r$ consecutive rows from a given tensor channel form a packet, as illustrated in Fig.~\ref{fig:tensor_packets}. Let $p_{i}^{(j)}$ be the set of $(x,y)$ coordinates of the $i$-th packet in the $j$-th tensor channel. We will interchangeably refer to $p_i^{(j)}$ as a \emph{packet} and the \emph{packet's location in the tensor}.

\label{sec:propoon sed}

In order to minimize the impact of packet loss, we employ an importance estimator to estimate the importance of each element in $\chi$. The importance estimator produces a tensor $\mathcal{I}$ of the same dimension as $\chi$, where each element of $\mathcal{I}$ is an estimate of the importance of the corresponding element in $\chi$. Using $\mathcal{I}$, FEC is assigned to the feature packets that contain the most important features, and then the feature packets and FEC packets are sent to the cloud. 
Upon reception, the lost feature packets are recovered to within the erasure correction capability of the FEC code~\cite{Lin_Costello_2004}. The data that could not be recovered is replaced by zeros. Finally, the resulting feature tensor is fed to the DNN back-end to produce the inference output. Our DNN backbone is ResNet-50~\cite{Resnet} split at layer \texttt{conv2\_block1\_add}, but the overall approach is applicable to other DNNs and split points.  

\subsection{Feature importance estimator}
The crucial problem in applying ULP to feature transmission is to decide on the relative importance of features. In our case of packet-based transmission, we will have to decide on packet importance. We present two methods for doing so: modified Grad-CAM and proxy model. 

\noindent\textbf{Modified Grad-CAM:} 
The original Grad-CAM~\cite{Grad-CAM} produces an importance estimate for each pixel in the input image. Hence, we make a few modifications to enable it to estimate feature importance, rather than input pixel importance.  
Specifically, we modify Grad-CAM by omitting globally averaging pooled gradients and summation of its channel importance maps. We extract such raw importance maps as a tensor $\mathcal{I}$ from the split-point of our DNN, such that it has the same dimension as the feature tensor $\chi$. Importance tensor $\mathcal{I}$ effectively provides an estimate of the importance of each feature in the feature tensor $\chi$. From $\mathcal{I}$, we compute the importance score $I_i^{(j)}$ of packet $p_i^{(j)}$ as: 
\begin{equation}
    {I_i}^{(j)}=\sum_{(x,y) \in p_{i}^{(j)}}\mathcal{I}(x,y,j).
    \label{eq:grad-cam_packet_importance}
\end{equation}
Then the values of $I_i^{(j)}$ can be used to decide which packets will be protected by FEC. While this approach provides a good estimate of the importance of each packet, as will be seen in the experiments, it is not realizable in practical CI. This is because Grad-CAM requires DNN's output, and passing gradients through DNN back-end, before $\mathcal{I}$ can be produced. Yet, in CI, we cannot obtain the DNN output before features are transmitted. Hence, we propose another method to estimate packet importance below.

\vspace{5pt}

\begin{figure}[t]
    \centering
    \includegraphics[width=0.49\textwidth]{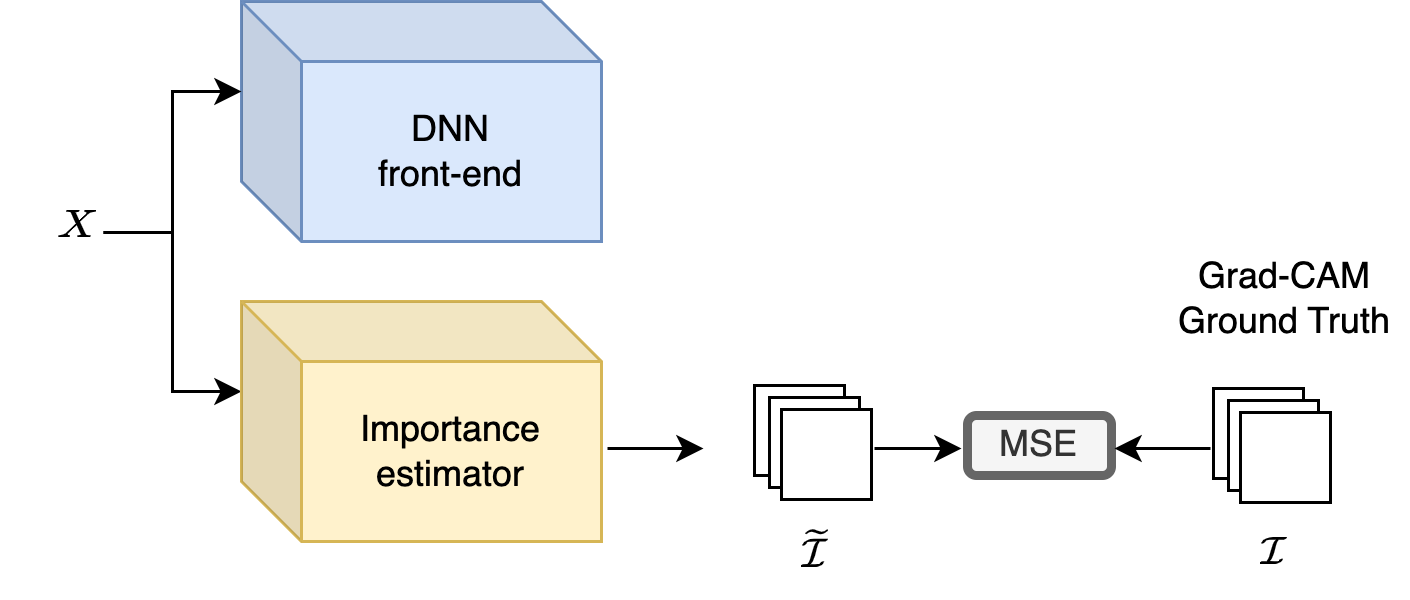}
    \caption{Proxy model training.}
   \label{fig:proxy}
\end{figure}

\noindent\textbf{Proxy model:} To enable estimating feature importance in a Grad-CAM-like manner, but without access to the DNN back-end or output, we propose to train a small DNN (``proxy model'') to estimates feature importance directly from the input. We take this proxy model to have the same architecture as the DNN front-end, in our case ResNet-50 up to layer \texttt{conv2\_block1\_add}. 
This proxy model is trained to approximate Grad-CAM's output. Specifically, let the proxy model's output be $\widetilde{\mathcal{I}}$, then the training loss function is $\mathcal{L}=$ MSE($\widetilde{\mathcal{I}},\mathcal{I}$), as shown in Fig.~\ref{fig:proxy}. Once the proxy model is trained, the packet importance can be computed based on its output as:
\begin{equation}
    {\widetilde{I}_i}^{(j)}=\sum_{(x,y) \in p_{i}^{(j)}} \widetilde{\mathcal{I}}(x,y,j).
    \label{eq:grad-cam_packet_importance}
\end{equation}

\begin{figure}[t]
    \centering
    \subfloat[\centering $\chi$]{{\includegraphics[width=0.16\textwidth]{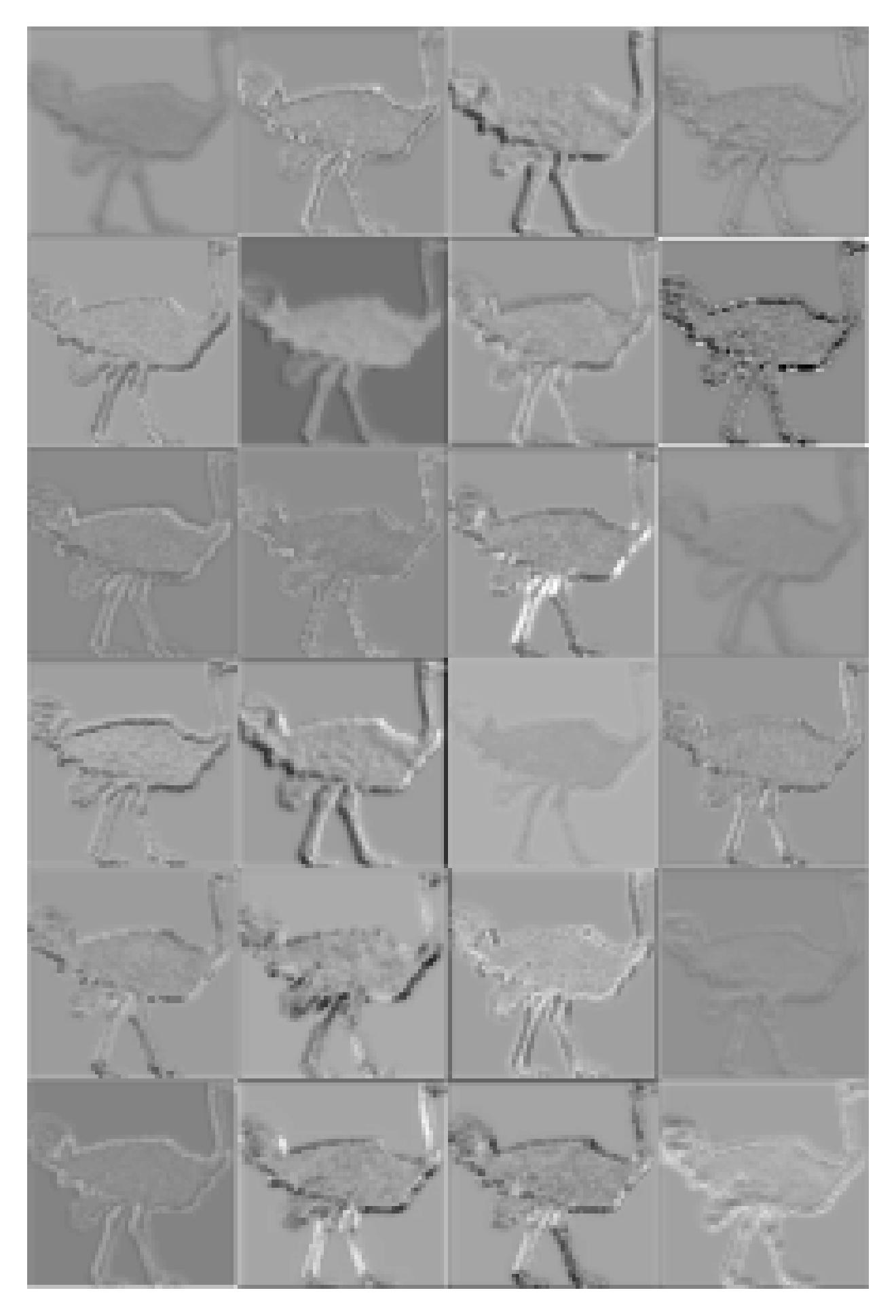} }}%
    \subfloat[\centering $I$]{{\includegraphics[width=0.16\textwidth]{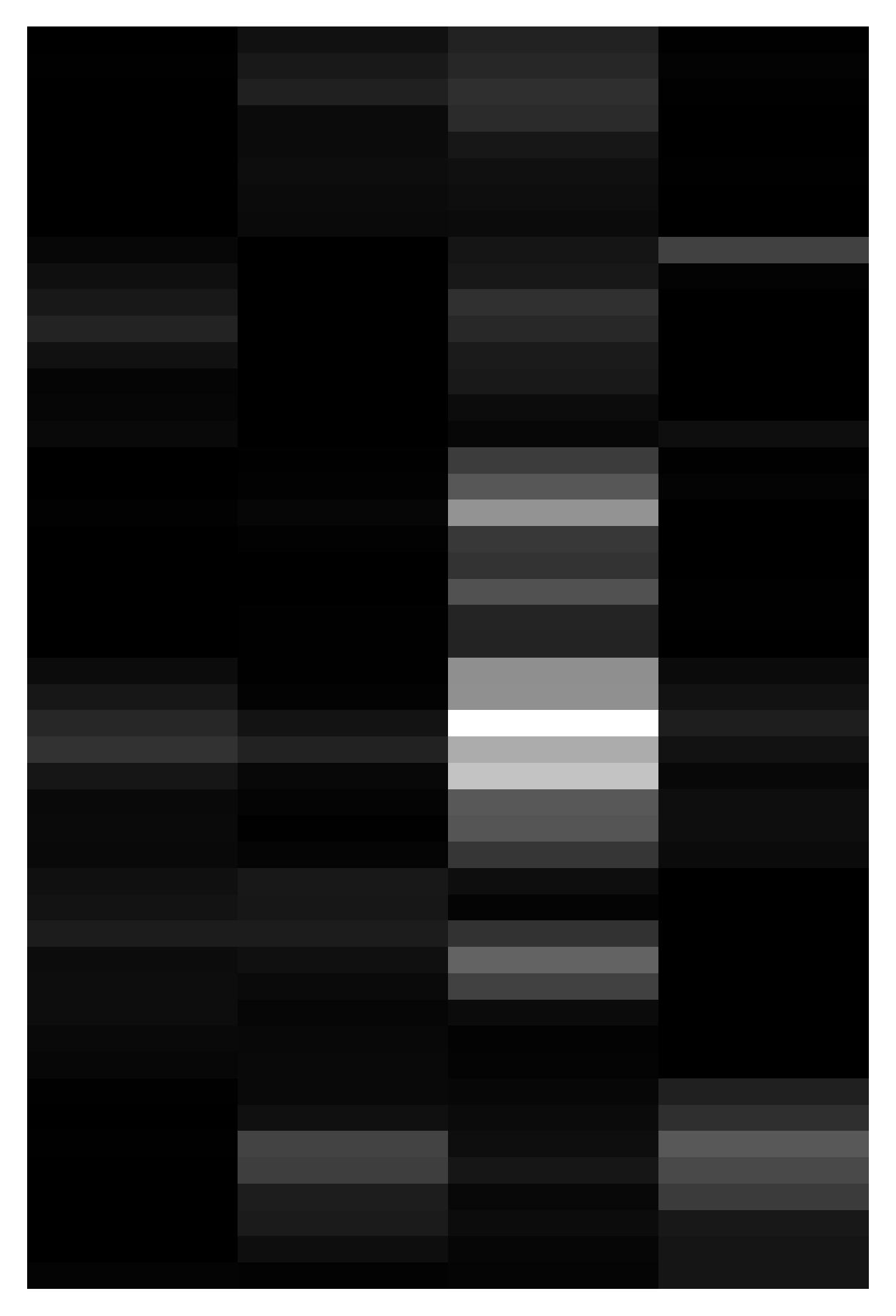} }}
    \subfloat[\centering $\tilde{I}$]{{\includegraphics[width=0.16\textwidth]{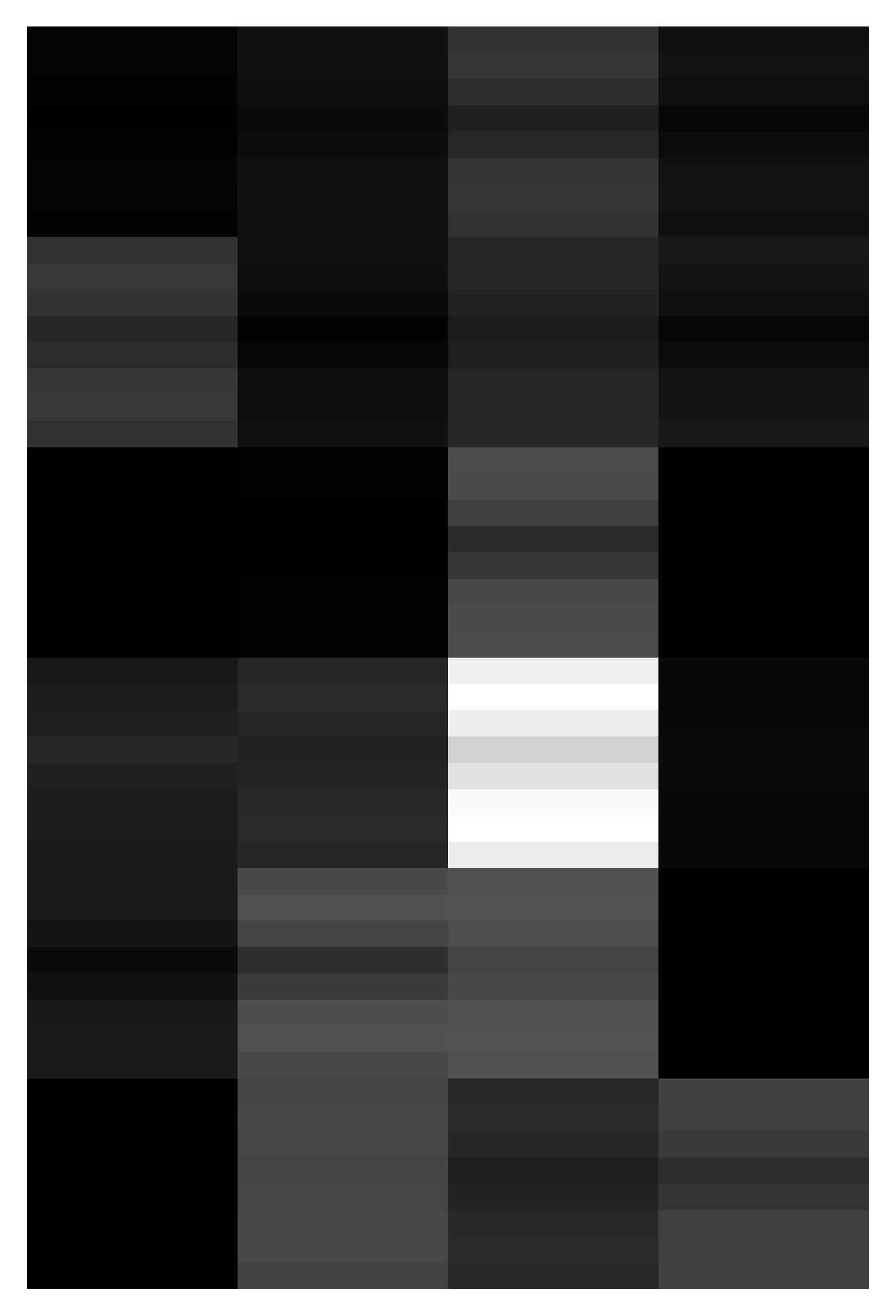} }}
    \caption{(a) Tiled feature tensor, and the packet importance obtained from (b) the modified Grad-CAM, and (c) the proxy model.}
    \label{fig:packet_importance}
\end{figure}

Fig.~\ref{fig:packet_importance}(a) shows a portion of the tiled feature tensor $\chi$ for an input image of an ostrich. Fig.~\ref{fig:packet_importance}(b) shows the corresponding packet importance $I_i^{(j)}$ produced by Grad-CAM, where each packet occupies $r=7$ rows of a feature tensor's channel. Fig.~\ref{fig:packet_importance}(c) shows the corresponding packet importance estimate $\widetilde{I}_i^{(j)}$ produced by the proxy model. While not equal to $I_i^{(j)}$, proxy's estimate correctly identifies the cluster of important packets. As will be seen in the results, the proxy model provides almost as good identification of important packets as Grad-CAM, while being fully realizable in CI.

\vspace{-10pt}
\subsection{Forward Error Correction}
We use systematic erasure correction Reed-Solomon (RS) codes~\cite{Rizzo_1997} as the Forward Error Correction (FEC) mechanism. Codewords of such codes are parametrized by two parameters, $(n,k)$, $n>k$, where $n$ is the total number of symbols and $k$ is the number of data symbols, so that $n-k$ is the number of redundancy symbols. Such codes have the property that the original $k$ data symbols can be recovered from any subset $k$ out of $n$ symbols; in other words, $n-k$ erasures can be corrected. By constructing codewords across packets~\cite{ULP_Mohr_JSAC2000}, one can create $n-k$ FEC packets for a group of $k$ data packets such that $n-k$ lost packets from the group of $n$ total packets can be recovered.  

To apply such FEC in our system, the packets of a given feature tensor are first sorted according to their estimated importance $\widetilde{I}_i^{(j)}$. Then a certain percentage of the important packets are protected by FEC. However, simply adding FEC packets would increase the amount of data to be transmitted. To address this issue, we adopt a strategy of dropping the least important packet for each added FEC packet. 
Specifically, A\% of the least important feature packets are removed and replaced by the same number of FEC packets, which then protect the remaining B\% of the feature packets. Obviously, A$+$B$=$100, and we refer to such a scheme as FEC\_A\_B.  This scheme provides unequal loss protection (ULP) because some feature packets are protected, some are not, and some are deliberately dropped in favor of FEC packets.

\section{Experiments}
\label{sec:experiments}
As mentioned earlier, we use the pre-trained ResNet-50 model~\cite{Resnet} split at layer \texttt{conv2\_block1\_add}, such that layers prior to this one form the DNN front-end and subsequent layers form the DNN back-end. To run the experiments, we utilized the same 882 images used in~\cite{CALTeC}, which are a subset of 10 classes of the ImageNet~\cite{imagenet2015} test set.
The tensor $\chi$ produced by the DNN front-end 
had dimensions of $56 \times 56 \times 256$ and was quantized to 8 bits per tensor element. The tensor was packetized into packets spanning $r=7$ rows, resulting in 8 packets per tensor channel. The proxy model was trained using the ImageNet~\cite{imagenet2015} training set, with 100 randomly selected samples from each of the 1000 classes. The Adam optimizer~\cite{adam_ICLR2015} was used, with the initial learning rate of $10^{-1}$ and decreasing to $10^{-11}$ over 16 epochs.

\begin{figure}[t]
    \centering
    \includegraphics[width=0.49\textwidth]{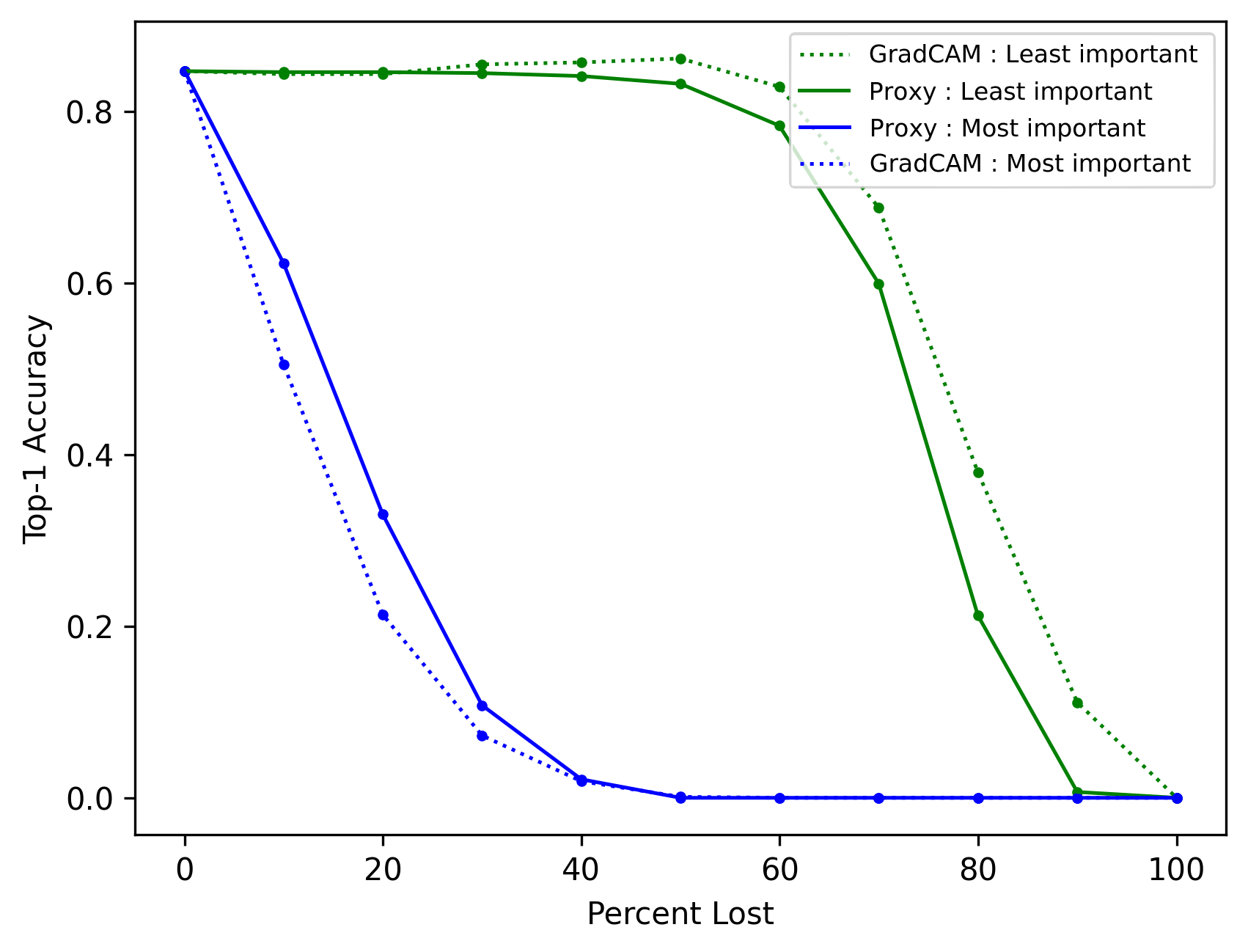}
    \caption{Top-1 classification accuracy vs. percentage of packets lost for the modified Grad-CAM and the proxy model.}
   \label{fig:Grad-CAM_vs_proxy}
\end{figure}

In the first experiment, we assess how well the modified Grad-CAM can identify important packets, and how well the proxy model can approximate Grad-CAM. The results are given in Fig.~\ref{fig:Grad-CAM_vs_proxy}, which shows the Top-1 classification accuracy as a function of the percentage of packets lost. The dashed curves correspond to importance estimates by the modified Grad-CAM: green curve for losing the least important packets and blue curve for losing the most important ones. Clearly, the modified Grad-CAM is able to separate important from non-important packets; when important packets are lost, the accuracy drops quickly, while when least important packets are lost, the accuracy is maintained even with 60\% lost packets. The proxy model, whose performance is indicated by solid curves, is not as good as the modified Grad-CAM, but is still able to identify important and less important packets reasonably well.

\begin{figure}[t]
    \centering
    \includegraphics[width=0.49\textwidth]{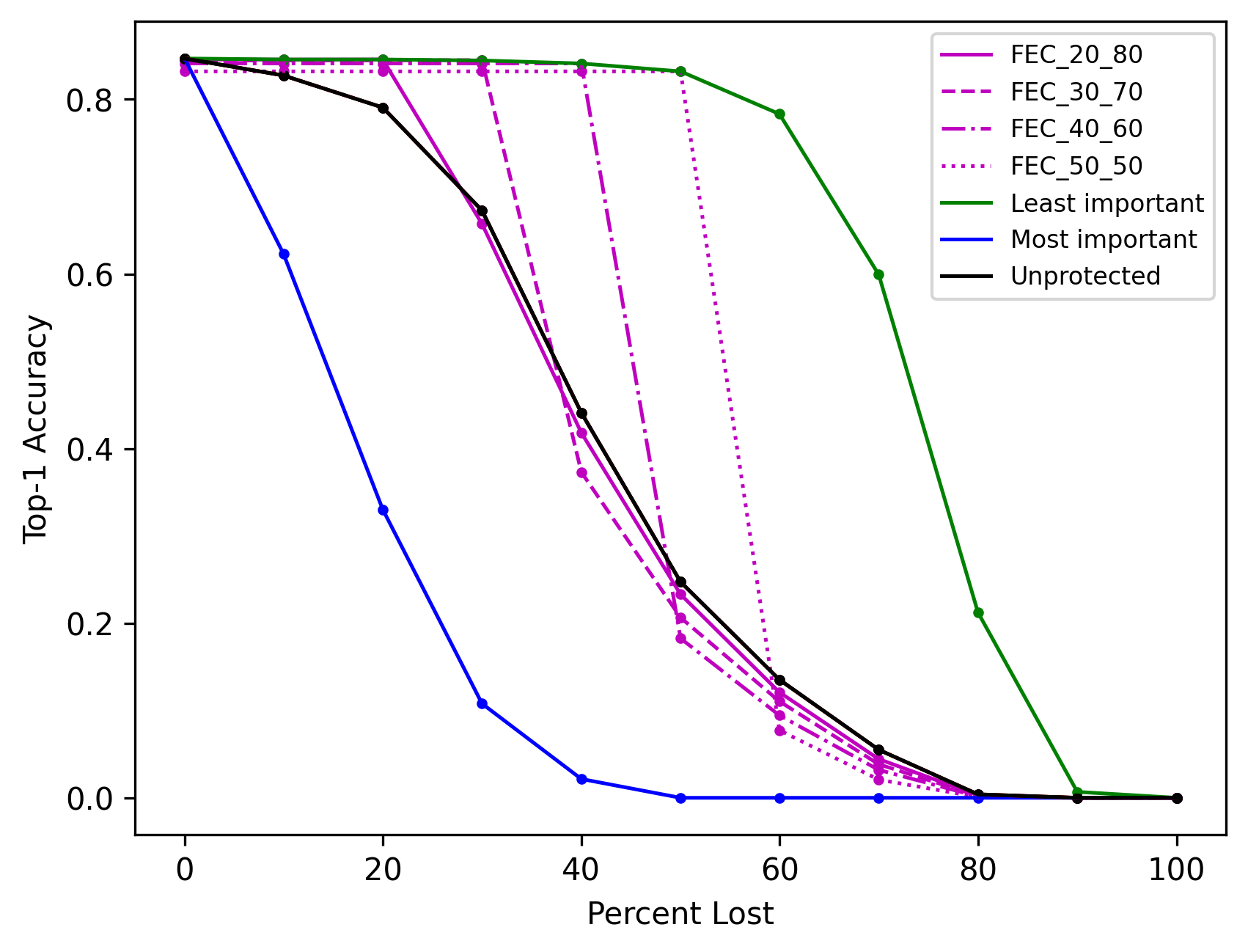}
    \caption{ Top-1 classification accuracy for ULP using the proxy model's importance estimates.}
    \label{fig:FEC}
\end{figure}

The next experiment is performed with an independent and identically distributed (iid) packet loss. Specifically, we choose a packet loss probability $P_L\in\{0.1, 0.2, ..., 1\}$ and run the experiment over the test set 10 times, recording the average Top-1 accuracy. Here, packet importance estimates are provided using the proxy model. In Fig.~\ref{fig:FEC} we show the results for several FEC\_A\_B schemes (pink curves) as well as unprotected packet transmission (red curve). We can see that as A increases, the performance improves because the increased amount of FEC protects the most important packets, while the least important feature packets are removed. Using FEC\_50\_50, we are able to achieve near-lossless performance even with 50\% packet loss. This also indicates that ResNet-50 produces fairly redundant features, otherwise maintaining the accuracy would not be possible while losing a large fraction of them. Our proposed method correctly identifies which of the features are important and ensures that those features are protected.

\section{Conclusions}
\label{sec:conclusion}

In this paper, we proposed a novel ULP-based approach that charts a new direction to improve the resilience of CI systems in the presence of packet loss. First, we demonstrated that a Grad-CAM-like approach is capable of providing reliable estimates of feature importance, which are needed for ULP. Second, we trained a proxy model for feature importance estimation, which can be deployed in a real CI system. The proxy model does not require access to the DNN output and makes its estimates solely based on the input image. Finally, we showed how the ULP mechanism and the feature importance estimator work together to prioritize the most important packets and minimize the impact of packet loss on the inference accuracy of the distributed DNN model. The resulting scheme significantly extends the range of packet loss rates over which reliable inference accuracy can be provided. 


\bibliographystyle{IEEEbib-abbrev} 
\small
\bibliography{ref}

\begin{thebibliography}{10}

\bibitem{kang2017neurosurgeon}
Y. Kang, J. Hauswald, C. Gao, A. Rovinski, T. Mudge, J. Mars, and L. Tang,
\newblock ``Neurosurgeon: Collaborative intelligence between the cloud and
  mobile edge,''
\newblock in {\em Proc. 22nd ACM Int. Conf. Arch. Support Programming Languages
  and Operating Syst.}, 2017, pp. 615--629.

\bibitem{CI_overview_ICASSP2021}
I.~V. Bajić, W. Lin, and Y. Tian,
\newblock ``Collaborative intelligence: Challenges and opportunities,''
\newblock in {\em Proc. IEEE ICASSP}, 2021, pp. 8493--8497.

\bibitem{intelligent_sensing_TIP2020}
Z. Chen, K. Fan, S. Wang, L. Duan, W. Lin, and A.~C. Kot,
\newblock ``Toward intelligent sensing: Intermediate deep feature
  compression,''
\newblock {\em IEEE Transactions on Image Processing}, vol. 29, pp. 2230--2243,
  2020.

\bibitem{NJSCC_ICML2019}
K. Choi, K. Tatwawadi, A. Grover, T. Weissman, and S. Ermon,
\newblock ``Neural joint source-channel coding,''
\newblock in {\em Proc. ICML}, 2019.

\bibitem{BottleNet++}
J. Shao and J. Zhang,
\newblock ``{BottleNet++}: An end-to-end approach for feature compression in
  device-edge co-inference systems,''
\newblock in {\em Proc. IEEE ICC Workshops}, 2020, pp. 1--6.

\bibitem{ALTeC}
L. Bragilevsky and I.~V. Bajić,
\newblock ``Tensor completion methods for collaborative intelligence,''
\newblock {\em IEEE Access}, vol. 8, pp. 41162--41174, 2020.

\bibitem{inpainting_ICC2021}
I.~V. Bajić,
\newblock ``Latent space inpainting for loss-resilient collaborative object
  detection,''
\newblock in {\em Proc. IEEE ICC}, 2021, pp. 1--6.

\bibitem{CALTeC}
A. Dhondea, R.~A. Cohen, and I.~V. Bajić,
\newblock ``{CALTeC}: Content-adaptive linear tensor completion for
  collaborative intelligence,''
\newblock in {\em Proc. IEEE ICIP}, 2021.

\bibitem{Grad-CAM}
R.~R. Selvaraju, M. Cogswell, A. Das, R. Vedantam, D. Parikh, and D. Batra,
\newblock ``{Grad-CAM}: Visual explanations from deep networks via
  gradient-based localization,''
\newblock in {\em Proc. IEEE ICCV}, 2017, pp. 618--626.

\bibitem{Resnet}
K. He, X. Zhang, S. Ren, and J. Sun,
\newblock ``Deep residual learning for image recognition,''
\newblock in {\em Proc. IEEE CVPR}, 2016, pp. 770--778.

\bibitem{ULP_Mohr_JSAC2000}
A. Mohr, E. Riskin, and R. Ladner,
\newblock ``Unequal loss protection: graceful degradation of image quality over
  packet erasure channels through forward error correction,''
\newblock {\em IEEE J. Selected Areas in Communications}, vol. 18, no. 6, pp.
  819--828, 2000.

\bibitem{ULP_vdS_TMM2001}
M. van~der Schaar and H. Radha,
\newblock ``Unequal packet loss resilience for fine-granular-scalability
  video,''
\newblock {\em IEEE Trans. Multimedia}, vol. 3, no. 4, pp. 381--394, 2001.

\bibitem{ULP_SPIC2003}
X. Yang, C. Zhu, Z. Li, X. Lin, G. Feng, S. Wu, and N. Ling,
\newblock ``Unequal loss protection for robust transmission of motion
  compensated video over the internet,''
\newblock {\em Signal Processing: Image Communication}, vol. 18, no. 3, pp.
  157--167, 2003.

\bibitem{Lin_Costello_2004}
S. Lin and D.~J. Costello,
\newblock {\em Error Control Coding},
\newblock Pearson, 2nd edition, 2004.

\bibitem{Rizzo_1997}
L. Rizzo,
\newblock ``Effective erasure codes for reliable computer communication
  protocols,''
\newblock {\em SIGCOMM Comput. Commun. Rev.}, vol. 27, no. 2, pp. 24–36, Apr.
  1997 1997.

\bibitem{imagenet2015}
O. Russakovsky, J. Deng, H. Su, J. Krause, S. Satheesh, S. Ma, Z. Huang, A.
  Karpathy, A. Khosla, M. Bernstein, A.~C. Berg, and L. Fei-Fei,
\newblock ``Imagenet large scale visual recognition challenge,''
\newblock {\em Int. J. Comput. Vision}, vol. 115, no. 3, pp. 211--252, Dec.
  2015.

\bibitem{adam_ICLR2015}
D.~P. Kingma and J. Ba,
\newblock ``Adam: A method for stochastic optimization,''
\newblock in {\em Proc. ICLR}, 2015.

\end{thebibliography}

\end{document}